# Rationality Problems of the Principles of Equivalence and General Relativity

## Mei   Xiaochun

(Department of Physics, Fuzhou University, E-mail: mxc001@163.com Tel:0086-591-87614214)
(N0.27-B, South Building, Zhongfu West Lake Garden, Fubin Road, Fuzhou, 350025, China)

**Abstract** It is pointed out that at present we only prove that inertial static mass and gravitational static mass are equivalent. We have not proved that inertial moving mass and gravitational moving mass are also equivalent. It is proved by the dynamic effect of special relativity that inertial moving mass and gravitational moving mass are not equivalent. Besides, it can be proved that the principle of general relativity is untenable. It is only an apparent felling of mankind actually. Therefore, there exists serious defect in the foundation of the Einstein's theory of gravitation. We need to renovate our basic idea of space-time and gravity.
**PACS Numbers:** 04.20.-q   03.30.+p   04.20.Cv   04.90.+e   04.20.Dw
            Inertial Moving Mass   Gravitational Moving Mass

1. **Equivalent principle is not consistent with special relativity**

We point out at first that the equivalent principle is not consistent with special relativity. Einstein used the reference frame of rotational desk to show the principle of equivalence and concluded that the space-time of gravitational field was curved. As shown I Fig. 2.1, suppose that the desk $K_1$ with radium $r$ and perimeter $2\pi r$ is at rest in beginning. Then let the desk turn around its center in a uniform angle speed $\omega$. According to the contraction of space-time contract in special relativity, observers on the resting reference frame of ground $K_0$ would find that the perimeter of desk becomes $2\pi r\sqrt{1-\omega^2 r^2/c^2}$ owing to desk's rotation, but the radium of desk would not contract for there is no motion in the direction of radium. Therefore, the radio of perimeter and radium would be less than $\pi$ and the observers on the ground would affirm that the space of rotating desk is non-Euclidean space.

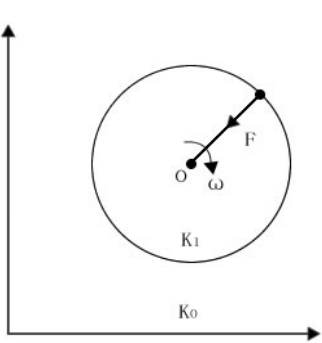 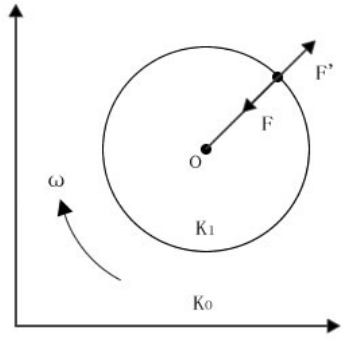 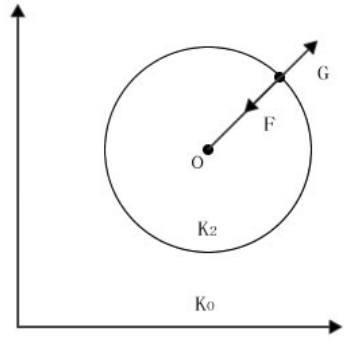

       Fig. 2.1                Fig. 2.2                Fig. 2.3

The problem now is what the observer resting on the rotating desk thinks about. Einstein and almost textbook of general relativity took a mistake here. It were though that because the speed of tangent direction caused length contraction, when these observers used their rulers to measure the perimeter of desk, the perimeter becomes $2\pi r/\sqrt{1-\omega^2 r^2/c^2}$, longer than that of resting desk. But when they use their rulers to measure the radium of desk, the length is unchanged. So the ratio of perimeter and radium is big than $\pi$. However, this result is wrong, for it is based on the premises that desk is at rest and observer and his ruler move. But in this case desk rotates actually so that the perimeter of desk and the rulers of observers contract synchronously. So the observers who are at rest on desk can not find the change of desk's perimeter actually by using their rulers.

Secondly, let's show that the principle of equivalence contradicts with the concept of space-time relativity. There are two problems here. The first is that according to the common understanding of special relativity, space-time's contraction is caused by relative speed, which has nothing to do with acceleration[1]. Or speaking strictly, the effect of acceleration is too small comparing with the effect of speed so that it can be neglected. So according to space-time relativity, for observers resting on rotating desk, $K_1$ is at rest but ground reference frame $K_0$ moves around $K_1$ shown in Fig. 2.2. In this way, the observers resting on desk would think that the rulers on their hands are unchanged so that they can not conclude that space-time of desk is curved. Instead, they would think that space-time of ground reference frame $K_0$ should be curved for $K_0$ is moving around $K_1$. Of course, this result can not be accepted. Therefore, if the concept of space-time relativity holds, we can not conclude that the space-time of gravitational field is curved. Conversely, if the space-time of gravitational field is curved and the principle of equivalence holds, the principle of space-time relativity would be violated.

The second problem is that in light of general recognition of special relativity, space-time contract is a purely relative effect, having nothing to do with acceleration. But according to the equivalent principle, space-time contraction should be relative to force and interaction, not a purely relative effect.

It is obvious that the principle of equivalence contradicts with the principle of space-time relativity. Because the principle of equivalence is regarded as the foundation of the Einstein's theory of gravitation, it can be said that special relativity is not consistent with the Einstein's theory of gravity. Only way for us to get rid of this paradox is to give up the relativity of space-time, and admit that space-time contraction is a kind of effect caused by accelerating processes. The rational result should be that the perimeter of rotating desk contracts and the radio of perimeter and radium is less than $\pi$. No matter who are on rotating desk or on ground, observer's viewpoints are the same.

## 2. Inertial moving mass and gravitational moving mass are nonequivalent

Let's discuss the problem of equivalent principle itself. The principle can be divided two parts, one is weak equivalent principle and another is strong equivalent principle. The weak equivalent principle indicates that gravitational mass is equivalent with inertial mass, or gravity is equivalent to inertial force locally. We fist discuss the equivalence of gravitational mass and inertial mass. Then discuss the equivalence of gravity and inertial force, as well as the problems of space-time contraction in gravitational field and non-inertial reference frame and so do.

At present, the experiments to prove the equivalence between inertial mass and gravitational mass is the so-called *Eötvös* type of experiments. It can be pointed out that this type of experiments can not expose the equivalence of inertial moving mass and gravitational moving mass actually. According to the second law of Newtonian, the force acting on an object is $F = m_{i0}a$. Here $m_{i0}$ is inertial static mass and $a$ is acceleration. In a uniform gravitational field, the formula of force can also be written as $F = m_{g0} g$. Here $m_{g0}$ is gravitational static mass and $g$ = constant is the strength of gravitational field. So we have

$$a = \frac{m_{g0}}{m_{i0}} g \qquad 2.1$$

If gravitational resting mass is equivalent with inertial resting mass, the radio $m_{g0} = m_{i0}$ should be a constant for any material. In this way, the acceleration of an object falling in this uniform gravitational field would also be a constant. This result has an obvious defect, i.e., the object's speed would surpass light's speed in vacuum at last when the object falls in the gravitational field. In order avoid this problem, the dynamic relation of special relativity should be considered. We have

$$m_i = \frac{m_{i0}}{\sqrt{1 - V^2/c^2}} \qquad F = \frac{d}{dt} \frac{m_{i0} V}{\sqrt{1 - V^2/c^2}} = \frac{m_{i0} a}{(1 - V^2/c^2)^{3/2}} \qquad 2.2$$

Because the effect of speed on gravitational force is still unknown at present, we suppose

$$m_g = f(V) m_{g0} \qquad 2.3$$

The function $f(V)$ is waited to be determined. Suppose we still have $F = m_g g$, Eq.(2.1) becomes

$$a = g f(V) \left(1 - \frac{V^2}{c^2}\right)^{3/2} \frac{m_{g0}}{m_{i0}} \qquad 2.4$$

As long as $f(V) \neq (1 - V^2/c^2)^{-3/2}$, the acceleration is not a constant again. When $V \to c$ we may have $a \to 0$, so that the falling speed would not surpass light's speed. It is obvious that when an object moves in a gravitational field, the effect of special relativity should be considered. In the *Eötvös* type of experiments, the moment acting on the hang bar is [2]

$$T = l_A a_T m_g^A \left( \frac{m_i^A}{m_g^A} - \frac{m_i^B}{m_g^B} \right) \qquad 2.5$$

Here $l_A$ is the length of bar, $a_T$ is the centrifugal acceleration on the level direction caused by the earth's rotation, $m_i^A$ and $m_i^B$ are the inertial masses, $m_g^A$ and $m_g^B$ are the gravitational masses of two different materials A and B individually. If the dynamic effect of special relativity is considered, by using formulas (2.2) and (2.3) Eq.(2.5) can be written as

$$T = \frac{l_A a_T m_{g0}^A}{\sqrt{1 - V^2/c^2}} \left( \frac{m_{i0}^A}{m_{g0}^A} - \frac{m_{i0}^B}{m_{g0}^B} \right) \qquad 2.6$$

It is obvious that $T$ has nothing to do with $f(V)$. As long as the radio $m_{i0}/m_{g0}$ does not change with different material, no matter what is the form of function $f(V)$, we always have $T = 0$. It means that the current experiments of the *Eötvös* types only verify the equivalence of gravitational static mass and inertial static mass without verifying the equivalence of gravitational moving mass and inertial moving mass. The relation between them will be deduced in later paper.

On the other hand, the weak principle of equivalence can also be described as that gravitation is locally equivalent with inertial force. Besides rotating desk, Einstein used the ideal experiment of closed chamber to show the equivalence. In the experiments, observers in a closed chamber could not decide whether they were at rest in a uniform gravitational field, or were being accelerated with a uniform acceleration. This result is correct in the Newtonian theory, but can be proved to be incorrect after the dynamic effect of relativity is considered. As shown in Fig. 2.4, suppose that the chamber $K'$ is accelerated relative to resting reference frame $K$. There is an object $A$ with static mass $m$ is hanged on the ceiling of chamber through an elastic rope. When $K'$ is accelerated in a uniform acceleration, owing to the dynamic effect of special relativity as shown in Eq.(2.2), the inertial mass of object would become bigger and bigger. So the rope would be pulled longer and longer besides the Lorentz contraction. When the speed of chamber is great enough, the inertial force of object would surpass the bearing capacity so that the rope would be pulled apart. Because the breaking of rope is an absolute event, observers who are inside and outside chamber would observe it. However, if observers were at rest in a static gravitational field, the rope would not be pulled longer and longer and break at last. The observers inside chamber would affirm that they are accelerated, instead of resting at a static gravitational field. It is obvious that the equivalence of gravitation and inertial force is only tenable in Newtonian mechanics. After the dynamic effect of special relativity is considered, gravitation and inertial force are not equivalent locally.

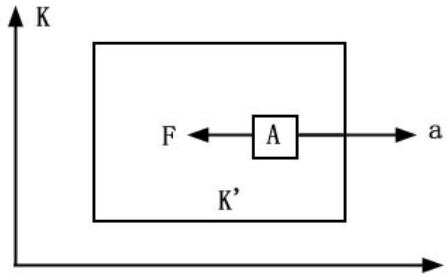

Fig. 2.4 Nonequivalence of gravitational moving mass and inertial moving mass

However, if the concrete form of function $f(V)$ is determined, under the promise that Eq.(3.4) is satisfied, we can still say that the gravitation and inertial force are still equivalent at a certain extent. The force acted on an object located on a rotating desk can be equal to a gravitation which does not change with time. The force acted on an object located in an accelerated and closed chamber can be equal to gravitation which changes with time. In fact, as we show later that gravitation is only a kind of special non-inertial force actually.

The equivalent problems of space-time contractions in gravitational fields and non-inertial reference frames are discussed below. Einstein used the ideal experiment of rotating desk to shown that the space-time of gravitational field was curved based on equivalent principle. We now discuss this problem. Because the clocks on rotating desk has a moving speed, its time would become slow with

$$\Delta t_1 = \Delta t_0 \sqrt{1 - \frac{r^2 \omega^2}{c^2}} \qquad 2.7$$

Here $\Delta t_1$ is the time of clocks on rotating desk, $\Delta t_0$ is the time of clocks on static ground. According to the principle of equivalence, as shown in Fig. 2.2., the reference frame of uniformly rotating desk is equivalent with a static gravitational field which does not change with time. In the figure, $F$ is the centripetal force acting on clock, $F'$ is inertial force. The equilibrium of two forces keeps the clock resting on the desk. According to the equivalent principle, as shown in Fig. 2.3, $F$ should be regarded as a certain force just as frictional force and so do. This force balances gravitation $G$ so that clock can be at rest in gravitational field without falling down. So as long as the equivalent principle is tenable, the time delay of clock which is at rest in gravitational field is expressed by Eq.(2.7). In the 60's last century, the experiments of atomic cesium clocks circling the earth had proved that clocks would become slow in gravitational field. They can be considered as the direct proofs that the existence of gravitational fields would cause time's delay. But there still exist a fundamental problem as shown below.

In atomic cesium clock, time is determined by atomic vibrations. The fact that atomic clock becomes slow in gravitational field indicates that the frequency of atomic vibration becomes slow, or the frequency of photon radiated by atom becomes slow and wave length becomes long or red shift. According to the theory of curved space-time, material's existence causes space-time curved. In general relativity, there is no the concept of force. Particles are not acted by force when they move along the geodesic line in curved space-time. Therefore, there is no the concept of potential energy for a photon moving in gravitational field. The total energy of a photon is equal to its dynamic energy. On the other hand, if the formula $E = h\nu$ is always tenable at any point of gravitational field and the frequency $\nu$ is not a constant, $E$ would change at different place in gravitational field. In this way, the law of energy conservation of photon would not be violated. However, the violation of energy conservation is unacceptable, though physicists seem to evade this problem at present. It will be shown in later discussion that we can establish gravitational theory in flat space-time. In the theory, photon would be acted by gravitation so the concept of potential energy is meaningful. Based on the theory, we can still reach the same result that gravitational field would cause time' delay without considering the principle of equivalence as a foundation for gravitational theory.

Now let's discuss the problem of space bending based on the equivalent principle. According to the current theory, when the angle speed of rotating desk is $\omega$, the space metric of desk is [3]

$$d\sigma^2 = dr^2 + \frac{r^2}{1 - r^2\omega^2/c^2}d\varphi^2 + dz^2 \qquad 2.8$$

It means that length contraction takes place along the tangent direction of desk. Because there is no speed along the radium direction, there exists no length contraction in this direction. Otherwise, according to the sphere symmetrical solution of the Einstein's equation of gravity, the space part of Schwarzschild metric can be written as

$$d\sigma^2 = \left(1 - \frac{\alpha}{r}\right)^{-1} dr^2 + r^2 \sin^2\theta \, d\varphi^2 + r^2 d\theta^2 \qquad 2.9$$

Because the slice of sphere is just the desk, so the situation of space bending should be the same. In fact, fore thin desk, we can let $dz = 0$ in Eq.(2.8). For the slice of sphere, we can let $\theta = \pi/2$ and $d\theta = 0$ in Eq. (2.9). In this case, Eqs.(2.8) and (2.9) should describe the same space metric. How, as we see that Eqs.(2.8) and (2.9) are not consistent in this case. In the Schwarzschild metric, length contraction takes place in the direction of radium, which is just the direction of gravitation. But in the metric of rotating desk Eq.(2.8), length contraction takes place in the tangent direction of desk, instead of the direction of inertial force. It indicates that space bending caused by non-inertial force may be different from that caused by gravitation, if gravitational fields would also cause space bending really. Unfortunately, this obvious difference has not caused person's attention up to now. It is obvious that even though the principle of equivalence holds so that gravity would cause space bending, the influences of non-inertial force and gravity on space may be different. In fact, it is lack of direct proofs to prove that gravity would cause space bending. What we have now is only indirect proofs. We only prove it by calculating object's motions in the gravitational field of the sun. The result is equal to space's bending at a certain degree. But they are indirect proofs and we need direct proofs. Even direct proofs prove that gravity would cause space bending in future, as shown in Eq.(2.9), the results may be different from that caused by non-inertial force. In this problem, further theoretical research and experiments are needed. In fact, in the theory of this paper, we can establish rational gravitational theory in flat space-time without the concept of curved space.

Besides, there are many other experiments to show the impossibility of weak equivalent principle. For example, when a closed chamber is at rest on a uniformly rotating desk, there exists the Coriolis force for a moving object, the inertial gyroscope would change its direction and charged objects would radiate photons. But in a static gravitational field, there are no such phenomena. Observers can distinguish both to be on rotating desk or at rest in a uniform gravitational field by theses phenomena. The problem of weak equivalence is very complex. Sometime it is effective, sometime it is wrong. Sometime it contains something specious. So it is improper to take it as a fundamental principle of physics.

## 3. Impossibility of the principles of strong equivalence and general relativity

The strong equivalent principle can be described as that for any space-time point in a gravitational field by taking the local inertial reference frame, the forms of nature laws are the same with that when the Descartes reference frame is taken without gravitational field and acceleration. Or speaking directly, in a local reference frame which is falling freely in a gravitational field, gravitation would be eliminated. This seems to be a well-known experimental fact. Then, is the strong equivalent principle tenable? The following analysis shows that this is only an apparent feeling of mankind and is wrong actually.

In order to explain it, we need to establish the concepts of "integral force" and "non-integral force". So-called "integral force" indicates that the force acted on an object is distributed uniformly over all parts of the object, even over every atoms and elemental particles of the object. "Non-integral force" indicates that the force is only acted on a part of the object. It is obvious that gravitation is "integral force", but the force acted on an observer who is at rest in non-inertial reference frame is non-integral force. For example, when an observer stands on an accelerated train, the force acted on him actually acts on his feet and the feet draw the body of observer moving forward. The other parts of body, not being acted directly by the force, are still in internal state, so that the body moves backward. In this way, so-called inertial force is caused. In an accelerating elevator, the force is only acted on observer's feet directly. Then the force is contributed to whole body through feet. For observers on rotating desk, friction force between feet and desk is need so that the observer can stand on desk. The friction force is also acted on feet directly. All of these forces are non-integral forces. It is just owing to the action of this kind of non-inertial force, non-equilibrium or internal stress is caused in observer's body so that observer can feel the existence of accelerating motion. It should be pointed out that "integral force" is only a macro-concept. Such as friction force, the essence of "non-integral force" is electromagnetic force. From micro-angle, however, all interactions between micro-particles are integral.

It is clear that the force acted on an object in uniform gravitational field is integral one. The inertial force caused by non-inertial motion is also integral force in general. The problem is now that when a non-integral force is distributed over whole body of observer who is at rest in a non-inertial reference frame what would happen? As shown in Fig. 3.1, the centripetal force acted on the feet of observer who is at rest on rotating desk is friction force actually. If this friction force is distributed over each part of observer's body, the observer would not fell this force's action. We can see in Fig.3.2 from the angle of non-inertial reference frame, this uniformly distributive friction force would be uniformly counteracted by the uniformly distributive inertial force. In this case, the observer who is at rotating desk seems to be at rest in a static inertial reference frame without any force's action, though he is actually accelerated. He may also think that he is falling in a gravitational field, just as an astronaut travels in the orbit of cincturing earth.

So it is only a subjective feeling that observer was not acted by force when he fell freely in a uniform gravitational field. The true is that the force acted on observer's body is integral force which is distributed over observer's body uniformly. In this way, no internal stress is produced in observer's body so that he can not fell the action of gravitational force and the existence of acceleration. Physiology tells us that felling is caused by non-uniformity. No felling does not indicate no changing. If changing is uniform and slow, there would be no felling. If proper methods are used, surpassing subjective felling, the observer would find himself to be being accelerated when he falls freely in a uniform gravitational field. For example, a charged object would radiate photons when it falls freely in a gravitational field, but the object would not when it is at rest in inertial reference frame. So only by taking a charged object with himself, the observer would judge whether he is falling freely in a uniform gravitational field or at rest in inertial reference frame. Only by this fact, the principle of strong equivalence has been proved untenable. More important is that comparing with static inertial reference frame, moving ruler's length would contract and moving clock's time would become slow as well as moving object's mass would increase absolutely. The forms of nature laws are different in both the freely falling reference frames of gravitational fields and in the Descartes inertial reference frame.

Einstein denied the existence of absolutely resting reference frame by establishing special relativity, thought that all inertial reference frames were equivalent for the description of nature phenomena. After that, he introduced the principle of general relativity again to eliminate the superiority of inertial reference frame, thought that all reference frames with arbitrary moving forms were equivalent for the description of nature phenomena. In mathematics, the principle of general relativity can be described as that the basic forms of motion equations should be covariant, or the line element of arbitrary reference frames must satisfy relation $ds^2 = $ constant. This demand is rational, for it actually represents the general form of the invariability principle of light's speed under the condition of non-inertial transformations. But it does not mean that the forms of forces acted on objects are the same. As we known, in the general three dimension space, the motion equations can be transformed into other forms by introducing arbitrary coordinate transformations without bring any practical physical result, for there are no any force or acceleration is introduced. But in the four dimension space-time, the situation is completely different. In the four dimension space-time, if the transformation is linear just as the Lorenz transformation, velocity would be introduced. But if the transformations are arbitrary, accelerations or non-inertial forces would be introduced in motion equations. In this way, there still exists a superior reference frame actually, i.e., inertial reference frame, in which motion equation has simplest form without the existence of inertial force.

In the Einstein's theory of gravity, non-inertial motions are considered to be equal to gravitational fields. In this situation, many problems are caused. At present, based on the principle of general relativity that any reference frame is equivalent to describe a physical system, when the coordinate transformation is carried out for a solution of gravitational field's equation, it is thought that the solution with new form would still be effective to describe the original gravitational field. However, even according to the principle of equivalence, this is also impossible. If non-inertial reference frame is equivalent locally with gravitational field, a special non-inertial reference frame can be only equal to a special gravitational field, with the corresponding relation of one by one. So when a non-inertial reference frame is transformed into another one, it means that a gravitational field is transformed into another one with different nature. That is to say, because inertial forces are considered to be equal to gravities, the general transformation in the space-time of four dimensions would change gravitational fields. For a certain gravitational field, suppose we have obtained its solution by solving the Einstein's equation. If the coordinate transformation is carried out, the form of field's equation and the solution are changed. The result means that the original field has been transformed into new gravitational field, and new solution would not be original one. So a certain gravitational field can only correspond to a certain metric. Arbitrary coordinate transformation in the space-time of four dimensions is not allowed, unless in new reference frame, the same result can be obtained. However, this is impossible in general. For example, we can not re-calculate the perihelion precession of Mercury as well as other problems in the Lemaite and Kruskal coordinate systems and get the same result as we do in the Schwarzschild coordinate system. In fact, the reason that the energy-momentum tensors of gravitational field can not be defined well at present is actually owing to the existence of strong equivalent principle. According to the principle, we can always introduce same local inertial reference frames through coordinate transformations, in which gravitational fields would be eliminated and the energy-momentum tensors of gravitational fields also become zero. However, according to the law of energy-momentum conservation, a static system's energy and momentum should be constants. How can they be canceled only by coordinate transformations?

Therefore, it can be said clearly that no relativity and arbitrarily are allowed in the description of gravity. The description of gravitational theory demands absolution. The principle of general relativity is untenable. The demand that the forms of motion equations should be covariant in arbitrary reference frames is only a basic restriction for the transformations of motion equations. It does not indicate that the descriptions of physical processes are the same in arbitrary reference frames. In fact, only in inertial reference frames, the basic forms of motion equations can be the same. And as mentioned in the paper "Absoluteness of Velocity Produced by Accelerating Process and Absolute Space-time Theory with Variable Scales", only in the absolute reference frame, the descriptions of physical processes are the simplest and most symmetrical.

At last, we should point out that the free falling of reference frames in gravitational field is only general accelerating motion. The force causing acceleration is only special one---- gravity. The result is that relative to resting reference frame, in accelerating reference frame, length would contract and time would delay and mass would increase. All of those are the normal effects of special relativity. Therefore, as long as we know the interaction form between gravitational field and material, we can describe object's motions in gravitational fields by the dynamic method of special relativity. The description is in flat space-time. It is unnecessary to us to introduce the equation of gravitational fields in curved space-time. This problem will be discussed in next paper.

It is obvious that though the Einstein's theory of gravity has reached great success, there are many problems existing in its logical foundation and theoretical configuration. We should renovate our ideas about the essence of space-time and gravity.